\documentclass[11pt,a4paper,english]{article}
\usepackage[right= 0.75in, left= 0.75in, top=1.05in, bottom=1.15in]{geometry}
\usepackage[latin1]{inputenc}
\usepackage[T1]{fontenc}
\usepackage[english]{babel}
\usepackage{pgf,tikz}
\usetikzlibrary{decorations.pathmorphing,patterns}
\usetikzlibrary{shapes.geometric}
\usetikzlibrary{arrows}
\usetikzlibrary{snakes}
\usepackage[format=hang,font=small]{caption}
\usepackage{booktabs,bm,multirow,subfig}
\usepackage{enumitem}
\usepackage{amsmath,amsfonts,amssymb,amsthm}
\usepackage{graphicx}
\usepackage[affil-it,auth-sc]{authblk}
\usepackage{cite}
\usepackage{color}
\usepackage{siunitx}
\usepackage{gensymb}
\graphicspath{{Images/}}
\newcommand{\dert}[1]{\dot{#1}}						

\renewcommand{\phi}{\varphi}
\renewcommand{\theta}{\vartheta}
\renewcommand{\epsilon}{\varepsilon}

\newcommand{\ds}{\displaystyle}
\catcode`^=\active
\renewcommand^[1]{\ensuremath{\sp{\mathrm{#1}}}}
\catcode`_=\active
\renewcommand_[1]{\ensuremath{\sb{\mathrm{#1}}}}
\numberwithin{equation}{section}
\tikzset{dashdotted/.style={dash pattern=on .8pt off 3pt on 4pt off 3pt}}
\pgfdeclaresnake{substrate}{initial}
{
\state{initial}[width=10pt]
{
\pgfpathlineto{\pgfpoint{0pt}{0pt}}
\pgfpathlineto{\pgfpoint{2.5pt}{0pt}}
\pgfpathlineto{\pgfpoint{2.5pt}{2.5pt}}
\pgfpathlineto{\pgfpoint{7.5pt}{2.5pt}}
\pgfpathlineto{\pgfpoint{7.5pt}{0pt}}
\pgfpathlineto{\pgfpoint{10pt}{0pt}}
}
\state{final}
{
\pgfpathlineto{\pgfpoint{0pt}{0pt}}
}
} 
\author[1,\footnote{Corresponding author -- e-mail address:\,\texttt{giovanni.noselli@sissa.it} -- phone:\,+39\,040\,3787\,432}]{Giovanni Noselli}
\author[1,2]{Antonio DeSimone}
\affil[1]{\small{SISSA - International School for Advanced Studies, via Bonomea 265, 34136 Trieste, Italy.} \smallskip}
\affil[2]{\small{GSSI - Gran Sasso Science Institute, viale Francesco Crispi 7, 67100 L'Aquila, Italy.} \smallskip}
\title{A robotic crawler exploiting directional frictional  interactions: \\
experiments, numerics, and derivation of a reduced model}
\date{\today}

\begin{document}
\maketitle
\begin{abstract}
\noindent
We present experimental and numerical results for a model crawler which is able to extract net positional changes from reciprocal shape changes, i.e. \lq breathing-like' deformations, thanks to directional, frictional interactions with a textured solid substrate, mediated by flexible inclined feet. We also present a simple reduced model that captures the essential features of the kinematics and energetics of the gait, and compare its predictions with the results from experiments and from numerical simulations. 
\end{abstract}
%

%
%
%
%
%
\section{Introduction}\label{sec:introduction}

The mechanics of locomotion at small scales is receiving increasing attention in the recent literature. This is due both to the intrinsic interest in the detailed understanding of the locomotion strategies of small biological organisms \cite{IMA,mcneil} and in the interest in reproducing them in artificial, bio-inspired artefacts \cite{hirose,Dreyfus}.
Depending on whether self-propulsion forces arise from the mechanical interactions of the locomotor with a surrounding fluid, or with a solid substrate, motility occurs by either swimming or crawling, and these are the main motility modes at microscopic scales \cite{fletcher-theriot}. 

The study of crawling at microscopic scales is often motivated by the interest in cell motility and spreading \cite{alberts,pollard}. In another stream of research, limbless locomotion has attracted interest as a new paradigm for robotic locomotion in very rough and complex environments or  on uneven terrains \cite{hirose,menciassi,tanaka}. Besides the engineering interest in view of the possible technological applications (rescue robotics, industrial inspection, medical endoscopy), the comparison of the behaviour of biological organisms with robotic replicas is particularly fruitful in that it promotes a synthetic, functionalist view of biological and bio-inspired motility, in which the essential necessary ingredients can be identified, and the way they interact can be studied in detail \cite{mahaSnakes,shelleySnakes,goldmannSnakes}.

One of the themes that  emerges naturally from the studies above is the identification of the mechanisms of minimal complexity that are able to produce nonzero net displacements by exploiting periodic shape changes. In swimming micro-motility, when the size of the swimmer is sufficiently small and the induced flows are characterized by low Reynolds numbers (Re), net displacements can be obtained only through non-reciprocal shape changes. This  fact  is known as the Scallop Theorem in the micro-swimming literature \cite{purcell}.
A geometric view emphasizing how the key to self-propulsion for low Re swimmers resides in performing closed loops in the space of shapes is discussed in \cite{alouges-1,alouges-2,alouges-3,alouges-4,arroyo-1,arroyo-2}, pursuing ideas pioneered in \cite{shapere-wilczek}.

In fact, we have shown in recent work that many ideas emerged in the context of low Re swimmers can be useful also in the crawling setting \cite{desimone-tatone,desimone-guarnieri,noselli-desimone}. A question that is particularly intriguing is whether, by exploiting sufficiently nonlinear interactions with a substrate one may \emph{beat} the Scallop Theorem, and achieve non zero net displacement for a locomotor that can only change shape in a reciprocal manner, through breathing-like deformations. Such deformations are represented by an open curve in the space of shapes, traveled forward and backward. While such shape changes produce zero net displacements in low Re swimming, this is no longer the case for crawlers on solid substrates, provided that the frictional interactions with the substrate are \lq directional' \cite{mahadevan,hancock}. A detailed study of the motion of model crawlers exploiting distributed, directional interactions is contained in \cite{gidoni}. The goal of this paper is to show how such a system can be realized in practice, and how a simple model of the type studied in \cite{gidoni} can be adequate to analyse its behaviour, resolving in a satisfactory way both the kinematics of the gait and its energetics.

\begin{figure}[h]
\renewcommand{\figurename}{Fig.}
\centering
\includegraphics[width=14.75cm]{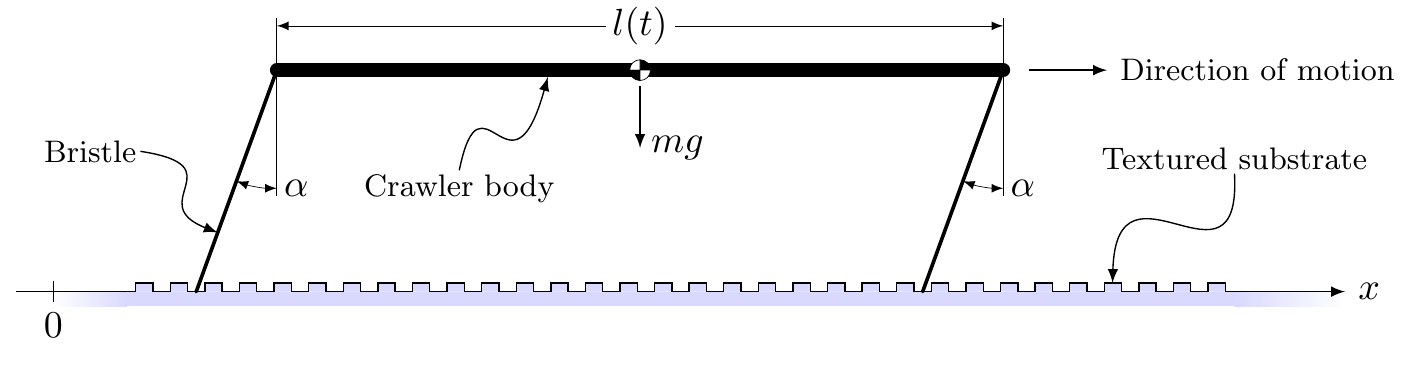}
\caption{A sketch of the model {\it bristle-crawler} analysed in this study. The system comprises an initially horizontal body of current length $l(t)$ and mass $m$, interacting with a groove-textured substrate by means of two inclined, flexible bristles. Due to the inclination of the elastic bristles, the frictional interaction arising at the crawler/substrate interface is directional in nature. Notice that the system is externally subject only to vertical gravitational loads.}
\label{fig:bristle-crawler}
\end{figure}

In this study, we shall consider the prototypical crawler  shown in Fig.~\ref{fig:bristle-crawler}. The system comprises an initially horizontal segment of current length $l(t)$ and mass $m$, interacting with a groove-textured surface by means of two inclined, flexible bristles. The inclination of the bristles by the angle $\alpha$ gives rise to directional frictional interactions, such that reciprocal length changes of the crawler body lead to its advancement on the substrate. 
Ours is only one of the possible choices to produce an effective frictional interaction with a force-velocity relation that is not odd. For example, one could texture the crawler instead of the substrate, and many alternative designs can be pursued based, for instance, on the bio-inspired surfaces reviewed in \cite{hancock}.

In the following, we will explore the crawler's motility both via direct experimentation and via finite element computations. These analyses will lead to the definition of a simplified, one-dimensional model capable of resolving both the kinematics and the energetics of the system being analysed. 

The focus of our paper is on the net displacements that can be extracted  from the most elementary form of cyclic shape changes, namely, reciprocal breathing-like deformations. Previous studies in the literature had mostly focused on peristaltic locomotion, where shape changes consist in traveling waves of extension or contraction \cite{mcneil,menciassi,tanaka}, a mechanism requiring a much more sophisticated level of spatio-temporal coordination. However, we share with these studies the interest in clarifying the basic aspects in the modelling of frictional interactions with a substrate, a necessary preliminary step on the way to obtain reliable models for the design of self-propelled micro-robots, bio-inspired by the crawling motility of invertebrate organisms.

\section{Experimental analysis of the \textbf{\textit{bristle-crawler}}}\label{sec:experiments}

\begin{figure}[h!]
\renewcommand{\figurename}{Fig.}
\centering
\includegraphics[width=15.5cm]{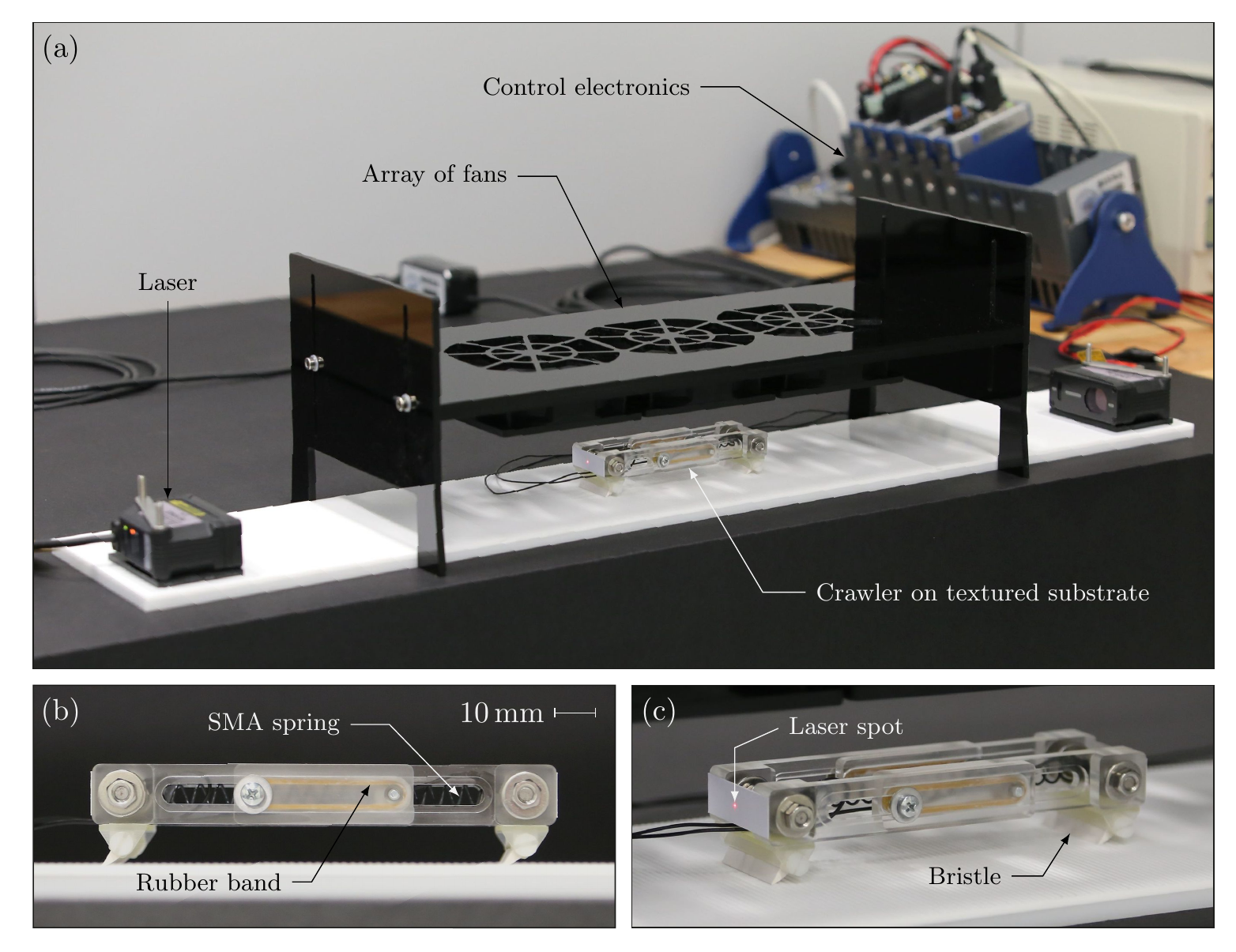}
\caption{The practical realization of the {\it bristle-crawler} analysed in this study. A global view of the experimental setting is reported in (a), whereas (b) and (c) are closer side views of the prototype crawler. Notice the SMA spring employed as actuator and the elastic rubber bands acting as antagonist, restoring elements. The length of the crawler, taken as the distance between the two bristles, is of 100\,mm for an overall mass of 55\,g, whereas the available shortening provided by the sliding mechanism is $\tilde s =$ 20\,mm.}
\label{fig:exp-1}
\end{figure}

In this section, we begin the analysis of the {\it bristle-crawler} sketched in Fig.~\ref{fig:bristle-crawler} by direct experimentation. The experimental setting employed in the study is shown in Fig.~\ref{fig:exp-1}. It comprises a prototype of the robotic crawler, positioned on a groove-textured substrate and actuated by means of a shape-memory-alloy (SMA) spring, an array of three fans, employed to enhance the convective cooling of the actuating coil, and the control/acquisition electronics.

All the components of the crawler's body were obtained using a EGX-600 CNC machine from Roland Corporation to engrave a 10\,mm thick plate of transparent PMMA, whereas the supports for the flexible bristles were realized with a 3DS Project 3510 HD printer to provide an inclination of $\alpha = \pi/4$\footnote{In the present study, we focus on the significance of directional interactions to crawling motility; a detailed analysis of the sensitivity of crawling performance to the number and inclination of the bristles is currently being performed  and will be the subject of a forthcoming paper.}. Similarly, a mold was 3D-printed for the preparation of the flexible bristles with a Shore A 40 silicone rubber. A tapered shape was chosen for the bristles in order to ensure interlocking between their tips and the substrate grooves: the cross section of $30\,\text{mm} \times 3\,\text{mm}$ at the bristle clamp linearly decreases to a cross section of $30\,\text{mm} \times 0.5\,\text{mm}$ at the bristle tip, for an overall length of 5\,mm.
Specifically, see Fig.~\ref{fig:exp-1}\,(b)-(c), the prototype crawler, of overall mass 55\,g and distance between the bristles 100\,mm, was designed symmetric about the anterior-posterior vertical plane, and each of its sides simply comprises two segments sliding side by side to provide a maximum shortening $\tilde s =$ 20\,mm via the resistive heating of the SMA actuator. The preparation of the textured substrate also required the use of the CNC machine to engrave evenly spaced rectangular symmetric grooves 1\,mm wide and 0.25\,mm deep in a 3\,mm thick plate of white PMMA.

For the actuation of the robotic crawler a SMA tension spring was chosen as a lightweight and compact solution. This was purchased from Jameco Electronics and comprises 20 active coils with an outer diameter of 6\,mm and a wire diameter of 0.75\,mm. Its connection to the crawler  body was achieved by means of two ending hooks. The shortening mechanism of the crawler was designed in such a way that its contraction implies the extension of two rubber bands, which act as antagonist, restoring elements for the SMA spring.

Each trial was carried out at room temperature (22\,\degree C) and with the three fans (from Jamicon Electronics, model JF0925-1UR) running at 3500\,rpm. The displacement at the left and right ends of the crawler  was measured with two triangulation, non-contact laser transducers (from Kyence, model IL-300) simultaneously sampled at a rate of 100\,S/s via a cRIO-9082 from National Instruments endowed with a NI-9215 AI module. This allowed for the determination of the crawler position, namely the displacement of its right hand side $\Delta(t)$, as a function of its shortening $s(t)$. The acquired shortening $s(t)$ was also exploited as feedback for a closed-loop control implemented on the FPGA of the cRIO-9082. In fact, at the beginning of each test an electric current of 3.5\,A was delivered to the SMA actuator through two tiny and flexible cables connected to a home-made amplifier, operating in commutation mode and controlled by a NI-9472 DO module, while monitoring the crawler shortening. When the shortening $s(t)$ attained the peak value of 20\,mm, the delivery of electric current was stopped by the control loop, allowing the robotic crawler to re-extend thanks the convective cooling provided by the fans. Again, the shortening $s(t)$ was monitored and no current was delivered to the actuator until the crawler returned to its original length. This cycle was repeated at least 5 times during each trial, and any individual stretching cycle approximately required a time of 20\,s.
\begin{figure}[h!]
\renewcommand{\figurename}{Fig.}
\centering
\includegraphics[width=12.5cm]{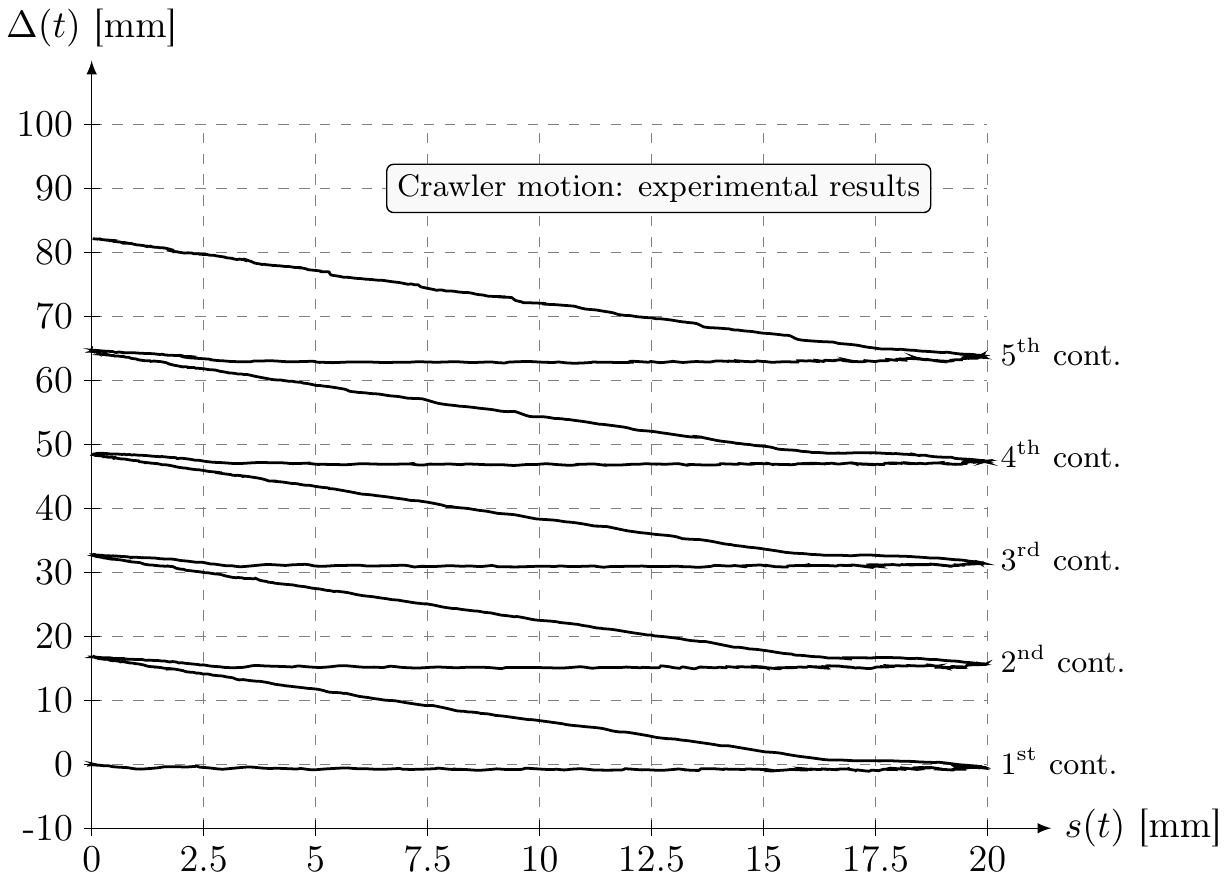}
\caption{Displacement of the right hand side of the crawler $\Delta(t)$ as a function of its shortening $0 \le s(t) \le \tilde s$. Five stretching cycles are reported, corresponding to a net advancement of approximately 82\,mm. Notice the changes in the slope of the diagram along the contraction and re-extension branches. These are due to the transitions between the {\it stick} and {\it slip} regimes of the two bristles. The advancement for each cycle is approximately $82\%$ of the available shortening of 20\,mm, a fact arising from the flexibility of the bristles.}
\label{fig:exp-2}
\end{figure}

The experimental analysis of the {\it bristle-crawler} allowed for the precise determination of its motile behaviour. During each trial we observed, as expected, a rightwards motion of the system, with the two bristles sliding at different times in the direction of less resistance (i.e., \lq along the grain' of the texture). This is also clearly visible from the movie provided in the electronic supplementary material (part I), which was taken with a digital camera EOS 6D from Canon, equipped with a EF 24-105\,mm 1:4 L IS USM objective. In particular, the movie shows the results of two distinct trials, carried out for an inclination $\alpha$ of the bristles of $\pi/4$ and 0, respectively. The second trial, with vertical bristles, was performed in order to experimentally prove the significance of the bristles inclination to directional interactions. In fact, for $\alpha = 0$, the motion of the system is characterized  by almost symmetric oscillations about its middle point, 
 so that no appreciable net advancement was observed during the test.

\begin{figure}[h!]
\renewcommand{\figurename}{Fig.}
\centering
\includegraphics[width=17.15cm]{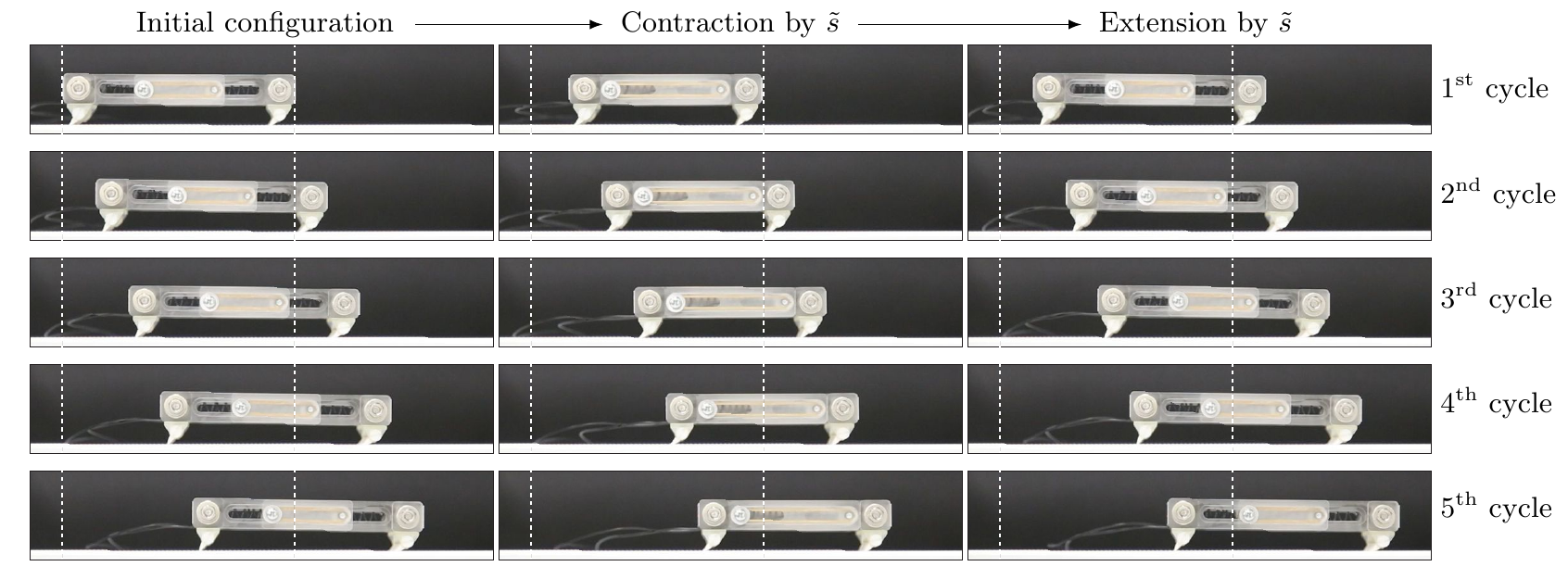}
\caption{Snapshots of the robotic crawler performing five stretching cycles. For each cycle, the initial configuration of the crawler is reported together with the two configurations after a contraction by $\tilde s$ and after the re-extension to the original length. Notice that vertical, dashed lines are drawn to highlight the displacement of the left and right extremities of the crawler's body. The last picture of each cycle is repeated for clarity as the first one of the following cycle.}
\label{fig:exp-3}
\end{figure}

The crawler's advancement, namely, the displacement $\Delta(t)$ of its right hand side, is shown in Fig.~\ref{fig:exp-2} as a function of the shortening $s(t)$. In particular, five stretching cycles are reported and these are highlighted in the figure. We notice that the net advancement arising from the five cycles approximately equals 82\,mm, and this is also the mean value measured from the experiments with a deviation of $\pm$ 1\,mm. Thus, the advancement arising from each stretching cycle is always smaller than the maximum available shortening of 20\,mm. The inspection of Fig.~\ref{fig:exp-2}, together with the snapshots in Fig.~\ref{fig:exp-3}, provide a rather clear explanation for that. In fact, during each cycle the crawler experiences two distinct regimes, related to a {\it stick-slip} behaviour of the bristles. During contraction, the two bristles initially deform elastically and their tips are at rest (the two bristles {\it stick}). Upon attainment of a critical shortening, the left bristle slides rightwards, whereas the other one keeps its position ({\it slip} of the left bristle and {\it stick} of the right one). Likewise, during re-extension the two bristles are first elastically deformed in the opposite direction with their tips at rest ({\it stick} of the two bristles), and then the right bristle slides rightwards, whereas the other one keeps again its position ({\it stick} of the left bristle and {\it slip} of the right one). It turns out that the flexibility of the two bristles which mediate the interaction with the substrate significantly affects the crawler motility. Indeed, a part of the available shortening is spent in each cycle, both during contraction and re-extension, to  bend elastically the bristles before sliding of their tips can take place.

The two regimes experienced by the crawler can be easily detected from the results of Fig.~\ref{fig:exp-2}, where the transition between {\it stick} and {\it slip} of the bristles typically corresponds to an evident change in the slope of the diagram along the contraction and re-extension branches. This is less visible during the first contraction due to the fact that, at the beginning of the test, the two bristles are almost unloaded and subject only to vertical, gravitational forces, whereas, at the beginning of all the following branches, the two bristles are always in a state of horizontal preloading.

Snapshots of the crawler  position during a typical test are reported in Fig.~\ref{fig:exp-3}. Specifically, five stretching cycles are shown for direct comparison with Fig.~\ref{fig:exp-2} and, for each cycle, the initial configuration of the crawler is reported together with the two configurations after a contraction by $\tilde s$ and after the re-extension to the original length. Notice that vertical, dashed lines are drawn to highlight the displacement of the left and right extremities of the crawler's body, whereas the last picture of each cycle is repeated for clarity as the first one of the following cycle. Inspection of the deformed shapes of Fig.~\ref{fig:exp-3} sheds light on the motile behaviour of the robotic crawler: the net advancement is achieved by a sequence of {\it stick-slip} events for the flexible bristles, with horizontal displacements much larger than the vertical ones. In the following, see Section~\ref{sec:one-dim}, we shall exploit these observations for the definition of an effective, one-dimensional model for crawling motility. 

\section{Finite element computations}\label{sec:fem}

The experimental analysis of the {\it bristle-crawler} allowed for the determination of the key mechanisms behind its kinematics. We extend now our study via numerical computations, by setting a finite element model capable of reproducing all the features of the crawler's motility that were observed during direct experimentation.

\begin{figure}[h!]
\renewcommand{\figurename}{Fig.}
\centering
\includegraphics[width=17.15cm]{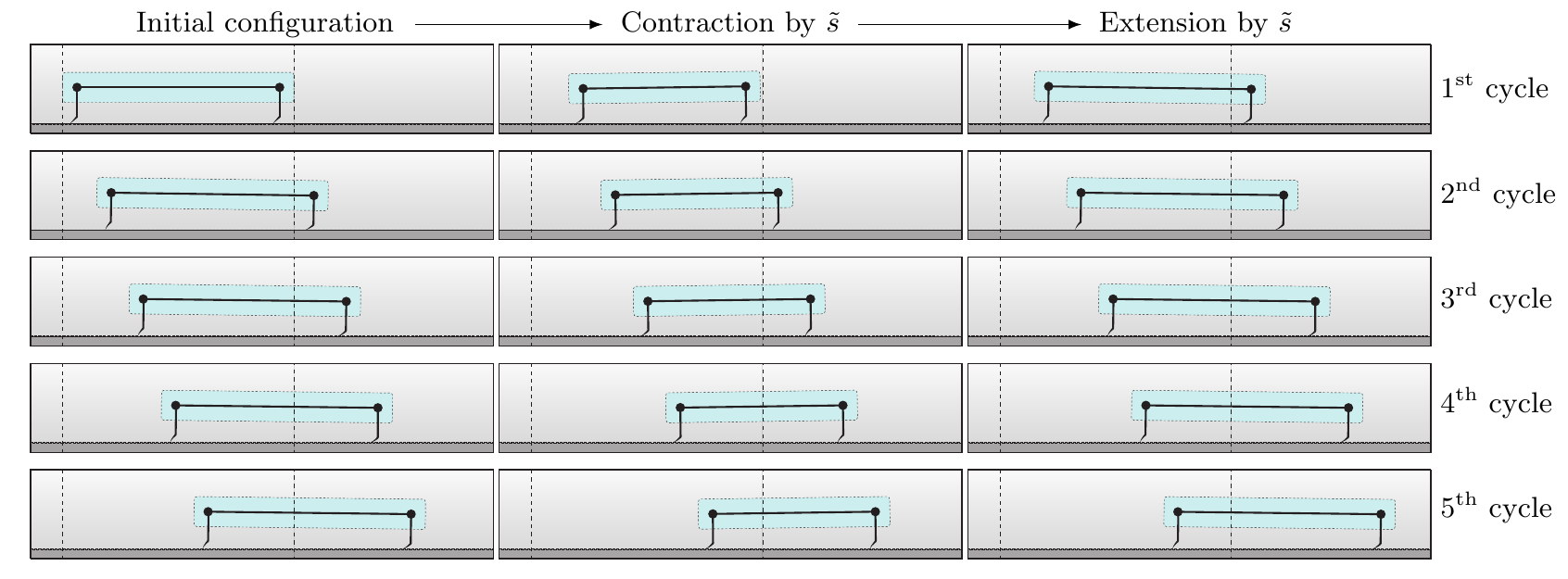}
\caption{Deformed shapes of the robotic crawler performing five stretching cycles. For each cycle, the initial configuration of the crawler is reported together with the two configurations after a contraction by $\tilde s$ and after the re-extension to the original length. Notice that vertical, dashed lines are drawn to highlight the displacement of the left and right hand side of the crawler's body. The last picture of each cycle is repeated for clarity as the first one of the following cycle.}
\label{fig:fem-2}
\end{figure}

The commercial software ABAQUS Standard 6.13-2 was employed to run the analyses. Specifically, the crawler  body was modeled as a horizontal connector of type CONN2D2 with initial length of 100\,mm, whereas the elastic bristles were modelled with two-nodes B21 beam elements. Their tapered shape was taken into account and, to this purpose, each bristle was discretized by means of hundred finite elements of variable cross section for an overall length of 5\,mm. With respect to the material properties, a Young's modulus of 1.1\,MPa, a Poisson's ratio of 0.45 and a mass density of 1.3\,g/cm^3 were chosen for the Shore A 40 silicone rubber, and viscous dissipation was accounted for by means of Rayleigh mass-proportional damping corresponding to a damping-ratio of 0.15 at the first resonant frequency. The flexible bristles were fixed to the crawler's body by means of rigid, vertical links ending at the extremities of the horizontal connector, where two point masses of 27.5\,g each 
 were also attached to account for the overall inertia of the system (these are depicted  as black spots in Fig.~\ref{fig:fem-2}, where also the  discretization  described above can be recognized). The textured substrate was introduced in the model as an analytical rigid surface, resembling the precise geometry employed in the experiments, and Coulomb dry friction was assumed at the contact between the bristles and the rigid surface. Remarkable agreement between the numerical predictions and the experimental results was obtained for a friction coefficient of 0.65, consistently with the properties of the two materials in  contact.

Each finite element analysis comprised two steps and both were run in a nonlinear, large deformation framework taking into account inertial effects. The crawler, positioned on the textured surface, was first subject to vertical gravitational loads (step 1), and then periodic shape changes were prescribed to the horizontal connector by means of a user subroutine DISP (step 2), with imposed shortening $s(t) = \tilde s \sin^2(\pi\,t/T)$, with a peak value $\tilde s = $ 20\,mm and a period $T = $ 20\,s.

\begin{figure}[h!]
\renewcommand{\figurename}{Fig.}
\centering
\includegraphics[width=12.5cm]{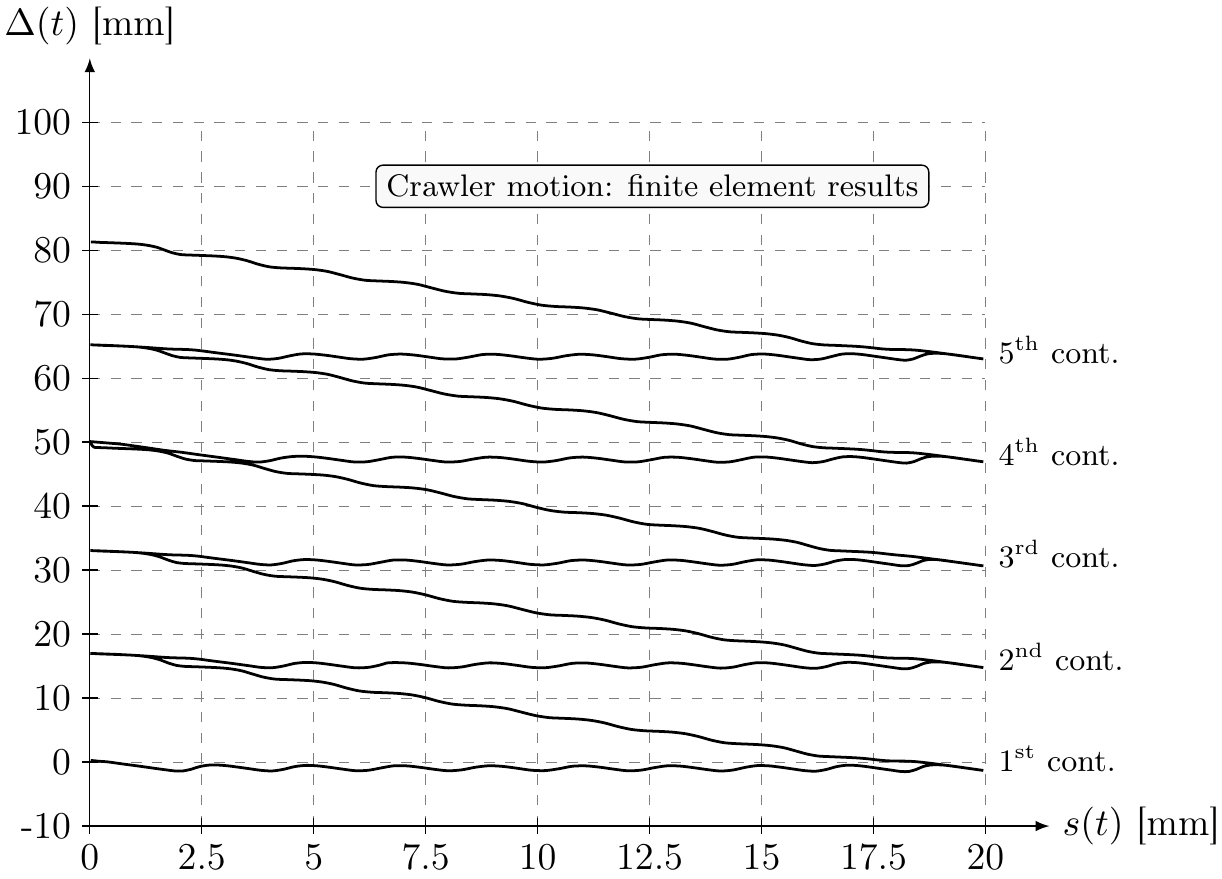}
\caption{Displacement of the right end of the crawler $\Delta(t)$ as a function of its shortening $0 \le s(t) \le \tilde s$. Five stretching cycles are reported, corresponding to a net advancement in line with the experimental measures, see Fig.~\ref{fig:exp-2}. Notice the small-amplitude oscillations in the advancement arising from the elasticity of the bristles and the discrete nature of the interaction between the bristles and the substrate. Just as in the experiments, the advancement in each cycle is smaller than the available shortening of 20\,mm, a fact arising from the flexibility of the bristles.}
\label{fig:fem-1}
\end{figure}

As a first numerical result, we show in Fig.~\ref{fig:fem-1} the crawler's advancement $\Delta(t)$ as a function of its shortening $s(t)$ for an inclination of the bristles  $\alpha = \pi/4$. In particular, five stretching cycles are reported, as highlighted in the figure, such that a direct comparison with the experimental results of Fig.~\ref{fig:exp-2} is feasible. A remarkable agreement was found for the set of material parameters reported above and, in fact, the finite element model well captures all the features of the crawler gait. Small-amplitude oscillations in the advancement $\Delta(t)$ are visible along the branches of the graph, and these arise from 
the elastic deformability of the bristles loaded by oscillating forces due to the discrete nature of the interaction between bristle tips and textured surface.

The finite element analyses were run accounting for the inertia of the system, but it seems reasonable, both from the experiments and the numerical computations, to conclude that for the prescribed time history of shape changes the crawler's gait falls within the quasi-static regime. We shall exploit  this observation in Section~\ref{sec:one-dim}.

Deformed shapes of the crawler are reported in Fig.~\ref{fig:fem-2}, and these were extracted from the numerical simulations. Again, five stretching cycles are reported and remarkable agreement is found with the experimental snapshots of Fig.~\ref{fig:exp-2}, a fact that further validates the adequacy of the proposed model to capture the gait of the crawler. The crawler's motion is also reported in the movie provided in the electronic supplementary material (part II), which shows the results from two distinct computations as obtained for an inclination of the bristles of $\alpha = \pi/4$ and $\alpha = 0$.

The computational results reported in Fig.~\ref{fig:fem-1} and in Fig.~\ref{fig:fem-2} mainly concern the kinematics of the system under investigation. However, in this study we are also interested in exploring the energy cost related to crawling motility, and this is feasible by computing the forces experienced by the connector while performing periodic shape changes. 
\begin{figure}[h!]
\renewcommand{\figurename}{Fig.}
\centering
\includegraphics[width=12.5cm]{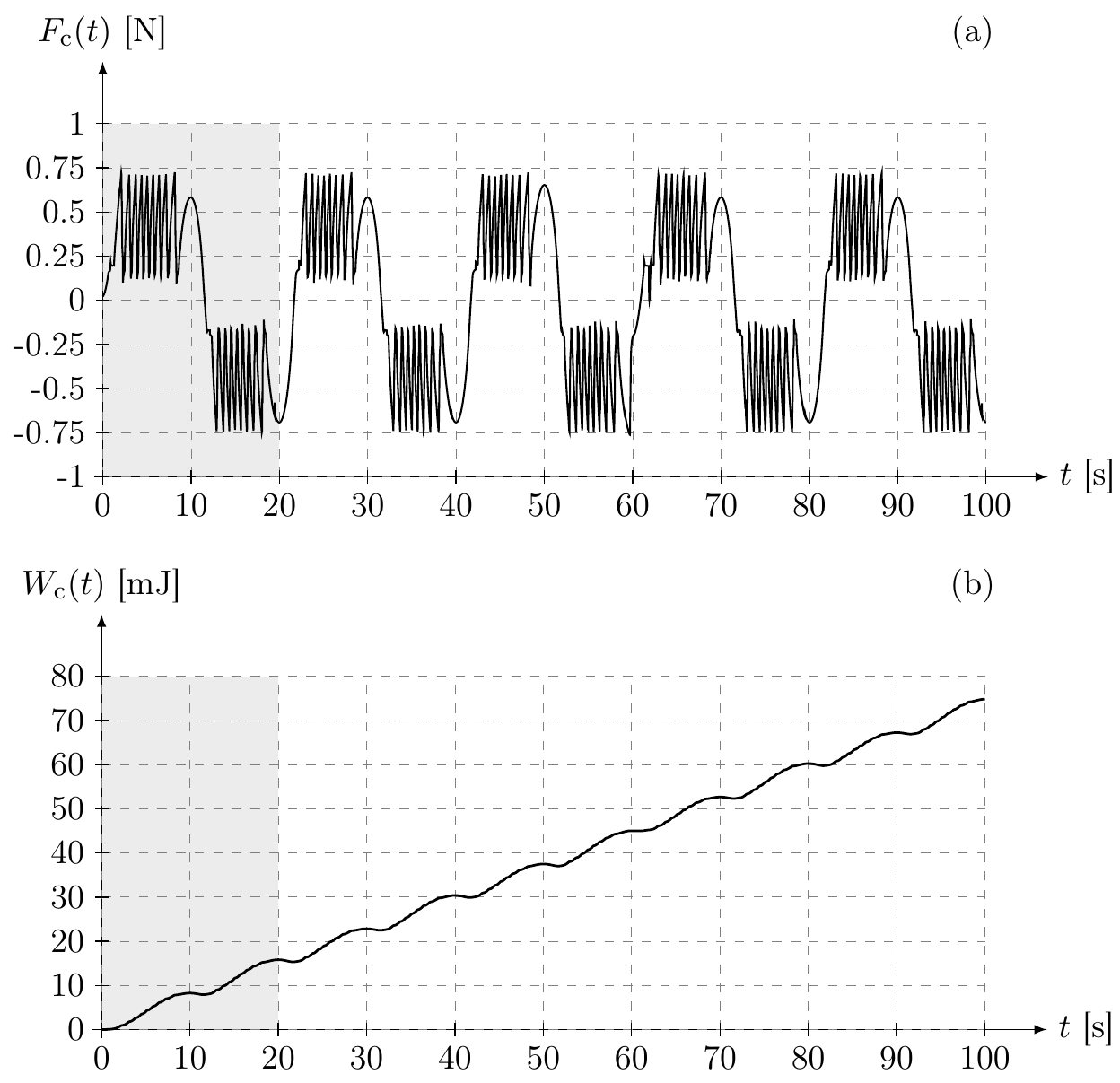}
\caption{Computed force $F_c(t)$ acting on the horizontal connector (a) and work performed by the connector $W_c(t)$ during cyclic shape changes (b). Notice that five cycles are reported, corresponding to a overall time of 100\,s for the analysis. The time interval for the first stretching cycle is highlighted in gray.}
\label{fig:fem-0}
\end{figure}
To this aim, results were extracted from the numerical computations and are summarized in Fig.~\ref{fig:fem-0} for the interesting case of $\alpha = \pi/4$. In particular, we report in Fig.~\ref{fig:fem-0}a the time history of the force $F_c(t)$ experienced by the axial connector during five stretching cycles, corresponding to an overall time of 100\,s. Likewise, the work $W_c(t)$ expended by the connector during those stretching cycles is shown in Fig.~\ref{fig:fem-0}b. The inspection of these results reveals the oscillatory, irregular nature of the frictional interaction arising at the crawler/substrate interface, and allows for a precise evaluation of the energetics associated with crawling motility in the context of directional interactions. As anticipated, we shall employ the experimental observations on the crawler gait and the computational results related to the energetics of crawling in the formulation of an effective, one-dimensional model.

\section{A one-dimensional model for the \textbf{\textit{bristle-crawler}}}\label{sec:one-dim}

In the previous sections, locomotion of the {\it bristle-crawler} has been extensively investigated both by means of experiments on a small-scale prototype and via detailed finite element computations. The analysis of those results, in particular of the deformed shapes shown in Fig.~\ref{fig:exp-3} and in Fig.~\ref{fig:fem-2}, reveals that the kinematics of the system essentially comprises horizontal displacements. In fact, the crawler  shortening $\tilde s$ much exceeds the vertical displacements, which only arise from the inflection of the bristles and from their adaptation to the irregularities of the substrate.

\begin{figure}[h]
\renewcommand{\figurename}{Fig.}
\centering
\includegraphics[width=10.25cm]{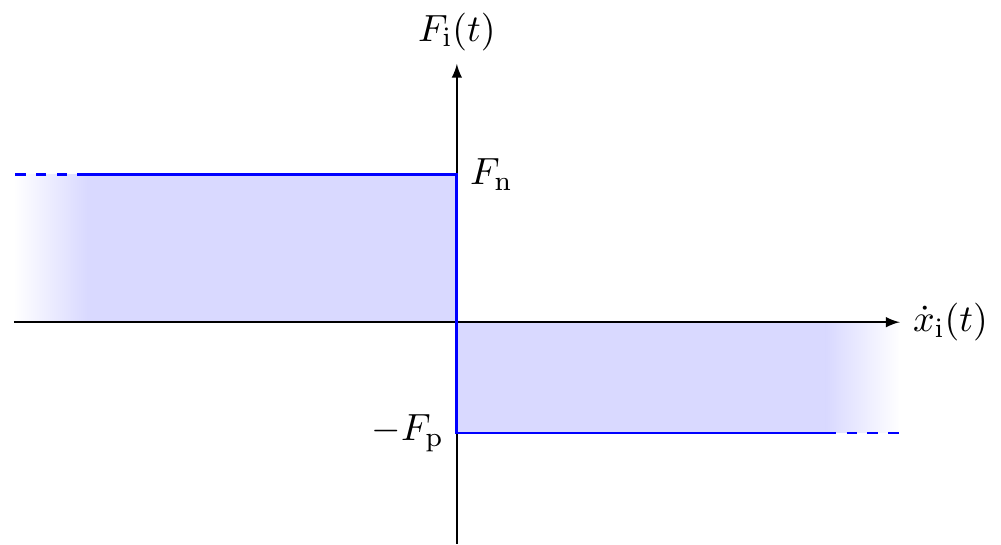}
\caption{A representation of the directional force-velocity law employed in this study. Due to the inclination of the bristles $F_n > F_p$, 
where the higher friction corresponds to negative sliding velocities.}
\label{fig:fric-law}
\end{figure}

In this study, directional interactions have been achieved through the sliding of inclined bristles on a groove-textured substrate. The question arises whether such interactions can be suitably modeled by means of an effective force-velocity law. 
In view of the results discussed in the previous sections,  a natural starting point is the following one-dimensional law:
\begin{equation}
F_i(t)=\begin{cases}
F_n					&	\text{if $\dert x_i(t)< 0$},\\
\tilde{F} \in [-F_p,F_n]	&	\text{if $\dert x_i(t)= 0$},\\
-F_p					&	\text{if $\dert x_i(t)> 0$},
\end{cases}
\label{eq:fric-law}
\end{equation}
where $F_i(t)$ is the horizontal force acting on the i-th bristle tip, function of its sliding velocity $\dert x_i(t)$ and of the non-negative parameters $F_p$ and $F_n$, see also Fig.~\ref{fig:fric-law}. In Eq.~\eqref{eq:fric-law}, the subscript $\mathrm{i}=\{1,2\}$, whereas the two parameters $F_p$ and $F_n$ correspond to the average frictional  forces acting on the bristle while sliding along the grain or against it, respectively. Notice that, due to the inclination of the two bristles by the angle $\alpha$, $F_n$ always exceeds $F_p$.

\begin{figure}[h!]
\renewcommand{\figurename}{Fig.}
\centering
\includegraphics[width=14.0cm]{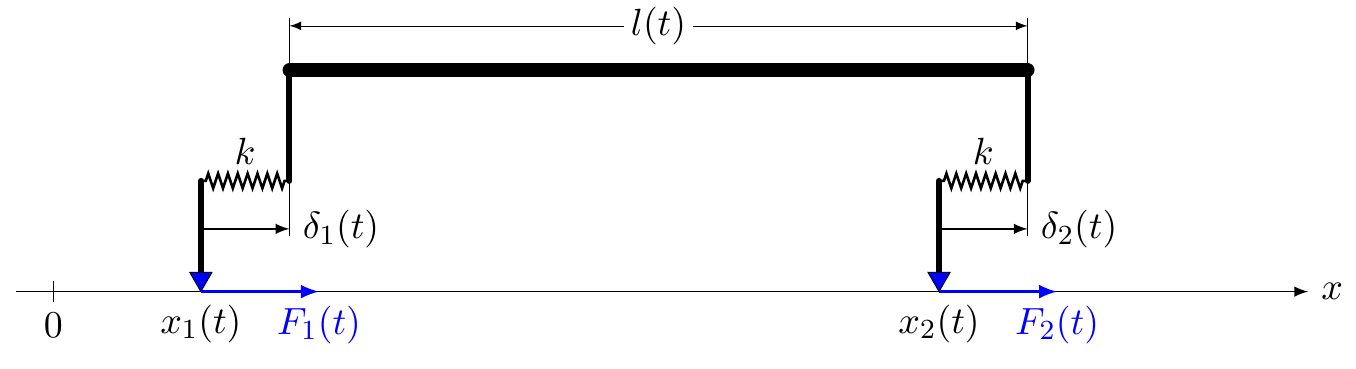}
\caption{A sketch of the one-dimensional model for the {\it bristle-crawler}. The model accounts only for horizontal displacement, and the two points $x_1(t)$ and $x_2(t)$ are subject to frictional, horizontal forces as given by the law of Eq.~\eqref{eq:fric-law}, see also Fig.~\ref{fig:fric-law}.}
\label{fig:one-dim-craw}
\end{figure}

It turns out that a simplified, one-dimensional model can be set for the bristle-crawler. This is shown in Fig.~\ref{fig:one-dim-craw}, where the crawler's body is depicted with a horizontal segment of current length $l(t)$.
The two bristles are modeled as vertical, rigid links and their compliance is accounted for by means of two linear springs of stiffness $k$ and unloaded length $\delta^0$, so that the tips of the bristles  (denoted by tiny solid triangles) are given an initial horizontal offset with respect to the crawler extremities. Frictional, directional interactions between the substrate and the bristles are described by  Eq.~\eqref{eq:fric-law}.

We extend now our approach to {\it quasi-static crawling} \cite{desimone-tatone,desimone-guarnieri,noselli-desimone,gidoni} by considering $T$-periodic, reciprocal shape changes of the system shown in Fig.~\ref{fig:one-dim-craw}, such that its length first monotonically decreases from the initial value of $L$ to $L- \tilde s$, and then monotonically increases from $L- \tilde s$ to $L$. At any instant $t$, the configuration of the system is known upon determination of the two co-ordinates $x_i(t)$ and of the two lengths $\delta_i(t)$, with $\mathrm{i}=\{1,2\}$. The compatibility of the displacements requires that 
\begin{equation}
x_2(t) + \delta_2(t) = x_1(t) + \delta_1(t) + l(t) ,
\label{eq:comp}
\end{equation}
whereas the balance of the horizontal forces acting on the crawler simply reads
\begin{equation}
F_1(t) + F_2(t) = 0 .
\label{eq:balance}
\end{equation}

Since the forces $F_i(t)$ are transmitted to the crawler body by the elastic springs, we can set
\begin{equation}
F_i(t) = -k[\delta_i(t) - \delta^0] , 
\label{eq:forces}
\end{equation}
and force balance provides the following relation between their lengths
\begin{equation}
\delta_2(t) = 2\delta^0 - \delta_1(t) . 
\label{eq:delta}
\end{equation}

The frictional law employed in this study requires the forces $F_i(t)$ to attain a constant yield value in order for sliding to occur, such that the problem under consideration possesses similarities with the constitutive theory for elastic, ideally-plastic solids \cite{lubliner} and can be conveniently approached in rate form. 
A similar approach has also been used to model the motion of capillary drops on rough substrates, see~\cite{gruenewald,hyst-droplets}.
Direct substitution of Eq.~\eqref{eq:delta} into Eq.~\eqref{eq:comp} and time differentiation yields 
\begin{equation}
\dert x_2(t) = \dert x_1(t) + 2 \dert \delta_1(t) + \dert l(t) ,
\label{eq:rate}
\end{equation}
an equation that can be integrated by splitting the stretching cycle in distinct stages. Furthermore notice that, in view of the restriction $F_n > F_p$, force balance dictates sliding to occur by a positive velocity and at one bristle tip at most, so that, while sliding, the crawler is subject to two balanced forces of modulus $F_p$.

\paragraph{\it Stage a:~$t \in [0,t_a)$ and $|F_i(t)| < F_p$.} We assume the crawler to be unloaded in its initial configuration, where $F_i(0) = 0$ and  $\delta_i(0) = \delta^0$. We further take $x_1(0) = 0$ and $x_2(0) = L$  as initial conditions. Hence, by setting $\dert x_1(t) = \dert x_2(t) = 0$ in Eq.~\eqref{eq:rate} we immediately get
\begin{equation}
\dert \delta_1(t) = -\frac{\dert l(t)}{2}  , 
\label{eq:der-delta-one}
\end{equation}
which, upon time integration and substitution into Eq.~\eqref{eq:forces}, provides the expressions of the two balanced forces as
\begin{equation}
F_1(t) = -F_2(t) = -\frac{k[L - l(t)]}{2} = -\frac{k\,s(t)}{2} ,
\label{eq:forces-a}
\end{equation}
where the shortening at time $t$ has been introduced as $s(t) = L - l(t)$. Notice that $F_1 < 0$ and $F_2 > 0$, so that Eq.~\eqref{eq:forces-a} holds as long as $F_1 > -F_p$. Therefore, at time $t = t_a$ the onset of sliding takes place at the tip of the left bristle for 
\begin{equation}
s(t_a) = \frac{2 F_p}{k} .
\label{eq:sliding-a}
\end{equation}

Upon definition of $\Delta(t) = x_2(t) + \delta_2(t) - L -\delta^0$, we obtain from the equations above the expression for the displacement of the right hand side of the crawler, namely
\begin{equation}
\Delta_a^1(t) = - \frac{s(t)}{2} \ge -\frac{F_p}{k} ,
\label{eq:Delta-a}
\end{equation}
with the superscript \lq 1' denoting the first stretching cycle.
It is worth noting that if $\tilde s < 2 F_p / k$ sliding cannot take place, so that, during re-extension, the system essentially recovers its original configuration by elastically unloading the two springs.

\paragraph{\it Stage b:~$t \in [t_a,t_b)$ and $|F_i(t)| = F_p$.} At the beginning of this second stage $s(t_a) = 2 F_p / k$, while the crawler is subject to balanced forces of modulus $F_p$. Hence, by further increasing the shortening up to $\tilde s$, the left bristle slides rightwards, whereas $\dert x_2(t) = \dert \delta_1(t) = \dert \delta_2(t) = 0$. Consequently, the displacement of the right hand side remains unchanged and reads as
\begin{equation}
\Delta_b^1(t) = \Delta_a^1(t_a) = -\frac{F_p}{k} .
\label{eq:Delta-b}
\end{equation}

\paragraph{\it Stage c:~$t \in [t_b,t_c)$ and $|F_i(t)| < F_p$.} Upon decreasing the shortening $s(t)$, the two elastic springs unload and the moduli of the forces, i.e. $|F_i(t)|$, decrease as well. Therefore, we can again set $\dert x_1(t) = \dert x_2(t) = 0$ in Eq.~\eqref{eq:rate}, so that, by time integration of Eq.~\eqref{eq:der-delta-one} and substitution into Eq.~\eqref{eq:forces}, we obtain the expressions of the forces as
\begin{equation}
F_1(t) = -F_2(t) = -\frac{k[s(t) - \tilde s]}{2} - F_p .
\label{eq:forces-c}
\end{equation}

As $s(t)$ decreases, the two forces first vanish, and then invert their signs. For sufficiently small values of $s(t)$, $F_1 > 0$ and $F_2 < 0$, so that Eq.~\eqref{eq:forces-c} holds as long as $F_2 > -F_p$. Hence, at time $t = t_c$ sliding takes place at the tip of the right bristle for 
\begin{equation}
s(t_c) = \tilde s - \frac{4 F_p}{k} .
\label{eq:sliding-a}
\end{equation}

By making use of the equations above and of the definition of $\Delta(t)$, we easily obtain the expression for the displacement of the right end of the crawler as
\begin{equation}
\Delta_c^1(t) =  -\frac{F_p}{k} + \frac{\tilde s - s(t)}{2} \le \frac{F_p}{k} .
\label{eq:Delta-c}
\end{equation}

\paragraph{\it Stage d:~$t \in [t_c,T]$ and $|F_i(t)| = F_p$.} At the beginning of the last stage $s(t_c) = \tilde s - 4 F_p / k$, while the crawler is again subject to balanced forces of modulus $F_p$. Hence, by decreasing the shortening to zero, the right bristle slides rightwards. By setting $\dert x_1(t) = \dert \delta_1(t) = \dert \delta_2(t) = 0$ into Eq.~\eqref{eq:rate} we get 
\begin{equation}
\dert x_2(t) = \dert l(t) ,
\end{equation}
which, upon time integration, provides the displacement of the right hand side as
\begin{equation}
\Delta_d^1(t) = \tilde s - s(t) - \frac{3 F_p}{k} .
\label{eq:Delta-d}
\end{equation}

The net advancement achieved during the first stretching cycle can be derived from Eq.~\eqref{eq:Delta-d} by setting $s(T) = s(0) = 0$, leading to
\begin{equation}
\Delta_{net}^1 = \tilde s - \frac{3 F_p}{k} .
\label{eq:net-adv}
\end{equation}

The result of Eq.~\eqref{eq:net-adv} concludes the analysis of the first stretching cycle and sheds light on the effect of the bristle stiffness on the crawler  motility. In fact, as already observed through direct experimentation and numerical computations, the net advancement is less than the available shortening, and their difference, i.e. $-3 F_p/k$, arises from the flexibility of the bristles. It is also worth noting that, due to the assumption of {\it quasi-static crawling}, the results above do not depend on the precise time history of the shortening $s(t)$, provided the contraction and re-extension of the crawler are monotonic in time.
\begin{figure}[t]
\renewcommand{\figurename}{Fig.}
\centering
\includegraphics[width=12.5cm]{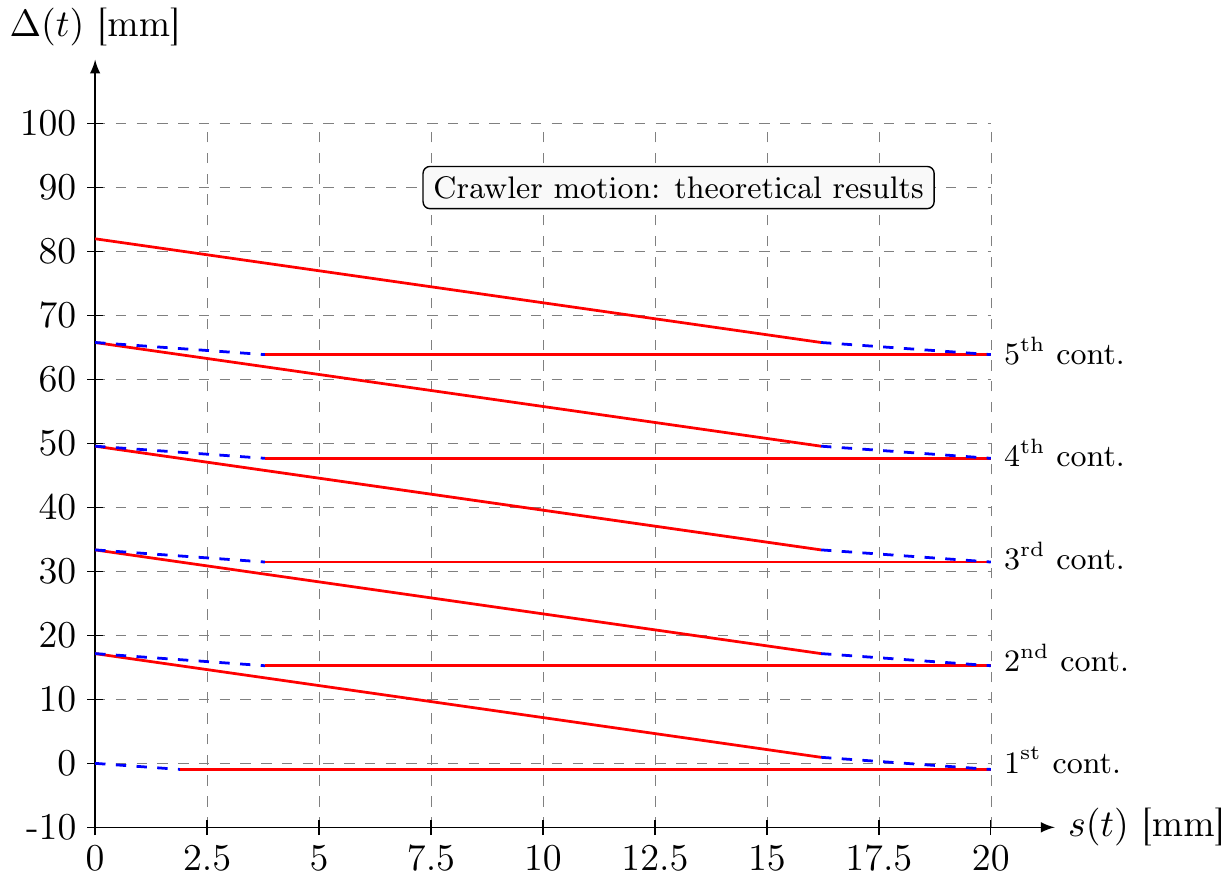}
\caption{Displacement of the right end of the crawler $\Delta(t)$ as a function of its shortening $0 \le s(t) \le \tilde s$. Notice that five stretching cycles are reported, with blue-dashed and red-solid segments denoting the two regimes of stick and slip for the bristles, respectively.}
\label{fig:res-one-dim}
\end{figure}

Subsequent stretching cycles can be similarly analysed, the only difference arising from the fact that, while at the beginning of the first cycle the crawler is unloaded, i.e. $F_i(0) = 0$, this is not the case of the following cycles and, in fact, $|F_i(T)| = F_p$.

Denoting by $n \ge 2$ the stretching cycle number, the expression for the displacement at the right hand side of the crawler follows as
\begin{equation}
\Delta^n(t)=\begin{cases}
\ds -\frac{s(t)}{2} + \Delta_{net}^1 + \left(\tilde s - \frac{4F_p}{k}\right)(n-2)							&	\text{for~ $\ds 0 \le s(t) < \frac{4 F_p}{k}$} ,\\[4mm]
\ds -\frac{2F_p}{k} + \Delta_{net}^1 + \left(\tilde s - \frac{4F_p}{k}\right)(n-2)						&	\text{for~ $\ds \frac{4 F_p}{k} \le s(t) \le \tilde s$} ,\\[4mm]
\ds -\frac{2F_p}{k} + \frac{\tilde s - s(t)}{2} + \Delta_{net}^1 + \left(\tilde s - \frac{4F_p}{k}\right)(n-2)		&	\text{for~ $\ds \tilde s - \frac{4 F_p}{k} < s(t) \le \tilde s$} ,\\[4mm]
\ds -\frac{4F_p}{k} + \tilde s - s(t) + \Delta_{net}^1 + \left(\tilde s - \frac{4F_p}{k}\right)(n-2)				&	\text{for~ $\ds 0 \le s(t) \le \tilde s - \frac{4 F_p}{k}$} .
\end{cases}
\label{eq:motion-sub}
\end{equation}

It turns out that the net advancement $\Delta_{net}^n$ for the subsequent stretching cycles is less than that of the first cycle by a factor $F_p/k$, namely
\begin{equation}
\Delta_{net}^n = \tilde s - \frac{4 F_p}{k} ,
\label{eq:net-adv-n}
\end{equation}
the difference arising again from the state of preload of the bristles at the beginning and at the end of each cycle such that $n \ge 2$.

The crawler motion is shown in Fig.~\ref{fig:res-one-dim} in terms of the displacement at its end $\Delta(t)$ as a function of the shortening $0 \le s(t) \le \tilde s$. The results were obtained for $\tilde s =$ 20\,mm and assuming $F_p/k =$ 0.95\,mm, so that direct comparison is possible both with the results of Fig.~\ref{fig:exp-2} and of Fig.~\ref{fig:fem-1}. Specifically, the value of $F_p/k$ was computed on the basis of the experimental and computational results in order to obtain from Eq.~\eqref{eq:motion-sub} an advancement of 82\,mm after five stretching cycles, i.e., for $n = 5$. Notice that the results of Fig.~\ref{fig:res-one-dim} well capture the features of the crawler  gait, such as the existence of the two regimes of stick and slip (and the related changes in the slope of the diagram along all the branches), shown as  blue-dashed and red-solid segments, respectively.

The analysis of the reduced one-dimensional model of Fig.~\ref{fig:one-dim-craw} is now completed by the evaluation of the work expended by the horizontal connector. For the first stretching cycle, this immediately follows from the Eqs.~\eqref{eq:der-delta-one}-\eqref{eq:net-adv}, namely
\begin{equation}
W^1 = \frac{F_p^2}{k} + 2 F_p\left(\tilde s - 3\frac{F_p}{k}\right) ,
\label{eq:work-first}
\end{equation}
and consists of a term of elastic energy stored in the bristles (first term) and a term of frictional dissipation (second term). Similarly, the work expended by the connector can also be computed for the following cycles, leading to
\begin{equation}
W^n = 2 F_p\left(\tilde s - 4\frac{F_p}{k}\right) ,
\label{eq:work-subs}
\end{equation}
an expression comprising only a frictional dissipation term and holding for $n \ge 2$. It is now worth noting that the expressions for the expended work only depend, at fixed gait, on the value of $F_p$ and on the ratio of $F_p/k$. As already discussed, kinematic arguments dictate the value of $F_p/k =$ 0.95\,mm, so that equating the work expended during five cycles as extracted from the finite element computations, namely 75\,mJ, with the corresponding expression from Eqs.~\eqref{eq:work-first}-\eqref{eq:work-subs}, i.e. for $n = 5$, a value of $F_p =$ 0.45\,N is obtained, and in turn $k =$ 0.48\,N/mm. The value obtained for the effective frictional force $F_p$  corresponds to the average force acting on the connector as computed via numerical simulations, see Fig.~\ref{fig:fem-0}a. Furthermore, the value of $k =$ 0.48\,N/mm is compatible with the horizontal, elastic stiffness of the flexible bristles as determined through beam theory, so that these values definitely confirm the adequacy of the proposed model in resolving both the kinematics and the energetics of the crawler being investigated.

\section{Conclusions and perspectives}

In this study, a model crawler capable to extract net positional changes from reciprocal, breathing-like deformations  was extensively analyzed. This remarkable ability of the system relies on directional  frictional interactions with a textured substrate, mediated by flexible inclined appendices. Both direct experimentation and non-linear finite element computations were exploited to extract the key features of the system, and a reduced, effective model was derived that well captures both the kinematics and the energetics of the crawler gait.

This simplified model provides us with a simple, yet powerful tool for the design and performance prediction of self-propelled robotic crawlers exploiting directional frictional interactions for locomotion.

Future work will consist in the study of the impact  of the  number and inclination of the bristles on the motility of the crawler. 
Moreover, different and more generic substrates will be considered, with the anticipation that smoother textures will produce less pronounced oscillations in the mechanical interactions, with smaller discrepancy between peak and average values of the interaction forces.

Finally, we plan to explore the possibility of actually implementing the concepts explored in this study in practical designs of smaller size by exploiting active materials such as Liquid Crystal Elastomers \cite{LCE-electric,LCE-bending}.  

\section*{Acknowledgments} We gratefully acknowledge the support by SISSA through the excellence program NOFYSAS 2012 and by the European Research Council through the ERC Advanced Grant 340685-MicroMotility.


\begin{thebibliography}{2}

\setlength{\itemsep}{-0.25mm}

{\small

\bibitem{IMA} Childress, S., Hosoi, A., Schultz, W.W. and Wang, J. 2012 {\it Natural locomotion in fluids and on surfaces: swimming, flying, and sliding}.
The IMA Volumes in Mathematics and its Applications No. 155. Springer Verlag, New York.

\bibitem{mcneil} McNeil Alexander, R. 2003 {\it Principles of Animal Locomotion}.
Princeton University Press, Princeton.

\bibitem{hirose} Hirose, S. 1993 {\it Biologically Inspired Robots: Snake-Like Locomotors and Manipulators}.
Oxford University Press, Oxford.

\bibitem{Dreyfus} Dreyfus, R., Baudry, J., Roper, M.L., Fermigier, M., Stone, H.A. and Bibette, J. 2005 Microscopic artificial swimmers.
{\it Nature} {\bf 437},  862-865. (doi:\,10.1038/nature04090)

\bibitem{fletcher-theriot} Fletcher, D.A. and Theriot, J.A. 2004 An introduction to cell motility for the physical scientist.
{\it Phys. Biol.} {\bf 1}, 1-10. (doi:\,10.1088/1478-3967/1/1/T01)

\bibitem{alberts} Alberts, B., Johnson, A., Lewis, J., Raff, M., Roberts, K. and Walter, P. 2002. {\it Molecular Biology of the Cell}.
Garland Science, New York.

\bibitem{pollard} Pollard, T.D and Earnshaw, W.C. 2008. {\it Cell Biology}.
Saunders, New York.

\bibitem{menciassi} Menciassi, A., Accoto, D., Gorini, S. and Dario, P. 2006. Development of a biomimetic miniature robotic crawler.
{\it Auton. Robot} {\bf 21}, 155-163. (doi:\,10.1007/s10514-006-7846-9)

\bibitem{tanaka} Tanaka, Y., Ito, K., Nakagaki, T. and Kobayashi, R. 2012 Mechanics of peristaltic locomotion and role of anchoring.
{\it J. R. Soc. Interface} {\bf 9}, 222-233. (doi:\,10.1098/rsif.2011.0339)

\bibitem{mahaSnakes} Guo, Z.V. and Mahadevan, L. 2008 Limbless undulatory propulsion on land.
{\it Proc. Nat. Acad. Sci. USA} {\bf 105}, 3179-3184 (doi:\,10.1073/pnas.0705442105).

\bibitem{shelleySnakes} Hu, D.L., Nirody, J., Scott, T. and Shelley, M.J. 2009 The mechanics of slithering locomotion.
{\it Proc. Nat. Acad. Sci. USA} {\bf 106}, 10081-10085 (doi:\,10.1073/pnas.0812533106).

\bibitem{goldmannSnakes} Maladen, R.D., Ding, Y., Umbanhowar, P.B., Kamor, A. and Goldman, D.I. 2011 Mechanical models of sandfish locomotion reveal principles of high performance subsurface sand-swimming.
{\it J. R. Soc. Interface} {\bf 8}, 1332-1345. (doi:\,10.1098/rsif.2010.0678)
 
\bibitem{purcell} Purcell, E.M. 1977 Life at low Reynolds number.
{\it Am. J. Phys.} {\bf 45}, 3-11. (doi:\,10.1119/1.10903)

\bibitem{alouges-1} Alouges, F., DeSimone, A. and Lefebvre, A. 2008 Optimal strokes for low Reynolds number swimmers: an example.
{\it J. Nonlinear Sci.} {\bf 18}, 277-302. (doi:\,10.1007/s00332-007-9013-7)

\bibitem{alouges-2} Alouges, F., DeSimone, A. and Lefebvre, A. 2009 Optimal strokes for axisymmetric microswimmers.
{\it Eur. Phys. J. E} {\bf 28}, 279-284. (doi:\,10.1140/epje/i2008-10406-4)

\bibitem{alouges-3} Alouges, F., DeSimone, A., Heltai, L., Lefebvre, A. and Merlet, B. 2013 Optimal swimming of Stokesian robots.
{\it Discrete Contin. Dyn. Syst. B} {\bf 18}, 1189-1215. (doi:\,10.3934/dcdsb.2013.18.1189)

\bibitem{alouges-4} Alouges, F., DeSimone, A., Giraldi, L. and Zoppello, M. 2013 Self-propulsion of slender micro-swimmers by curvature control: N-link swimmers.
{\it Int. J. Non-Linear Mech.} {\bf 56}, 132-141. (doi:\,10.1016/j.ijnonlinmec.2013.04.012)

\bibitem{arroyo-1} Arroyo, M., Heltai, L., Mill\'an, D. and DeSimone, A. 2012 Reverse engineering the euglenoid movement.
{\it Proc. Nat. Acad. Sci. USA} {\bf 109}, 17874-17879. (doi:\,10.1073/pnas.1213977109)

\bibitem{arroyo-2} Arroyo, M. and DeSimone, A. 2014 Shape control of active surfaces inspired by the movement of euglenids.
{\it J. Mech. Phys. Solids} {\bf 62}, 99-112. (doi:\,10.1016/j.jmps.2013.09.017)

\bibitem{shapere-wilczek} Shapere, A. and Wilczek, F. 1989 Geometry of self-propulsion at low Reynolds number.
{\it J. Fluid Mech.} {\bf 198}, 557-585. (doi:\,10.1017/S002211208900025X)

\bibitem{desimone-tatone} DeSimone, A. and Tatone, A. 2012 Crawling motility through the analysis of model locomotors: two case studies.
{\it Eur. Phys. J. E} {\bf 35}, 85. (doi:\,10.1140/epje/i2012-12085-x)

\bibitem{desimone-guarnieri} DeSimone, A., Guarnieri, F., Noselli, G. and Tatone, A. 2013 Crawlers in viscous environments: linear vs non-linear rheology.
{\it Int. J. Non-Linear Mech.} {\bf 56}, 142-147. (doi:\,10.1016/j.ijnonlinmec.2013.02.007)

\bibitem{noselli-desimone} Noselli, G., DeSimone, A. and Tatone, A. 2013 Discrete one-dimensional crawlers on viscous substrates: achievable net displacements and their energy cost.
{\it Mech. Res. Commun.} {\bf 58}, 73-81. (doi:\,10.1016/j.mechrescom.2013.10.023)

\bibitem{mahadevan} Mahadevan, L., Daniel, S. and Chaudhury, M.K. 2004 Biomimetic ratcheting motion of a soft, slender, sessile gel.
{\it Proc. Nat. Acad. Sci. USA} {\bf 101}, 23-26. (doi:\,10.1073/pnas.2637051100)

\bibitem{hancock} Hancock, M.J., Sekeroglu, K. and Demirel, M.C. 2012 Bioinspired directional surfaces for adhesion, wetting, and transport.
{\it Adv. Funct. Mater.} {\bf 22}, 2223-2234. (doi:\,10.1002/adfm.201103017)

\bibitem{gidoni} Gidoni, P., Noselli, G. and DeSimone, A. 2014. Crawling on directional surfaces.
{\it Int. J. Non-Linear Mech.}, {\bf 61}, 65-73. (doi:\,10.1016/j.ijnonlinmec.2014.01.012)

\bibitem{lubliner} Lubliner, J. 1990 {\it Plasticity theory}. Pearson Education.

\bibitem{gruenewald} DeSimone, A., Grunewald, N. and Otto, F. 2007 A new model for contact angle hysteresis.
{\it Netw. Heterog. Media} {\bf 2}, 211-225. (doi:\,10.3934/nhm.2007.2.211)

\bibitem{hyst-droplets} Fedeli, L., Turco, A. and DeSimone, A. 2011 Metastable equilibria of capillary drops on solid surfaces: a phase field approach.
{\it Continuum Mech. and Thermodyn.} {\bf 23}, 453-471. (doi:\,10.1007/s00161-011-0189-6)

\bibitem{LCE-electric} Fukunaga, A., Urayama, K., Takigawa, T., DeSimone, A. and Teresi, L. 2008 Dynamics of electro-opto-mechanical effects in swollen nematic elastomers.
{\it Macromolecules} {\bf 41}, 9389-9396. (doi:\,10.1021/ma801639j)

\bibitem{LCE-bending} Sawa, Y., Urayama, K., Takigawa, T., DeSimone, A. and Teresi, L. 2010 Thermally driven giant bending of LCE films with hybrid alignment.
{\it Macromolecules} {\bf 43}, 4362-4369. (doi:\,10.1021/ma1003979)


}

\end{thebibliography}
\end{document}